\documentclass[12pt]{article}
\usepackage{amsmath,amssymb,bm,graphicx}
\usepackage{color} 

\begin{document}

\begin{flushright}
\end{flushright}

\vskip 0.5 truecm

\begin{center}
{\Large{\bf  Topology change from a monopole to a dipole in  Berry's phase 
}}
\end{center}
\vskip .5 truecm
\centerline{\bf Shinichi Deguchi~$^1$
 {\rm and} Kazuo Fujikawa~$^2$ }
\vskip .4 truecm
\centerline {\it $^1$~Institute of Quantum Science, College of 
Science and Technology}
\centerline {\it Nihon University, Chiyoda-ku, Tokyo 101-8308, 
Japan}
\vskip 0.4 truecm
\centerline {\it $^2$~Interdisciplinary Theoretical and Mathematical Sciences Program, 
}
\centerline {\it   RIKEN, Wako 351-0198, 
Japan}

\vskip 0.5 truecm

\makeatletter
\makeatother

\begin{abstract}
The smooth topology change of Berry's phase from 
a Dirac monopole-like configuration to a dipole configuration, when one approaches the monopole position in the parameter space, is analyzed  in an exactly solvable model.  A novel aspect of  Berry's connection ${\cal A}_{k}$ is that the geometrical center of the monopole-like configuration and the origin of the Dirac string
are displaced in the parameter space. Gauss' theorem
$\int_{S}(\nabla\times {\cal A})\cdot d\vec{S}=\int_{V} \nabla\cdot (\nabla\times {\cal A}) dV=0$ 
for a volume $V$ which is free of singularities shows that a combination of the monopole-like configuration and the Dirac string is effectively a dipole. 
The smooth topology change from a dipole to a monopole with a quantized magnetic charge $e_{M}=2\pi\hbar$  takes place when one regards the Dirac string
as unobservable if it satisfies the Wu-Yang gauge invariance condition. In the transitional region from a dipole to a monopole, a half-monopole appears with an observable Dirac string, which is analogous to the  Aharonov-Bohm phase of an electron for the magnetic flux generated by the Cooper pair condensation.  
The main topological features of an exactly solvable model are shown to be supported by a generic model of Berry's phase.
\end{abstract}

 
\section{Introduction}
Berry's phase is defined for the level crossing
phenomenon~\cite{Higgins, Berry, Simon, Aharonov} and a monopole-like object~\cite{Dirac, Wu-Yang} appears at the level crossing point in the adiabatic approximation. The appearance of monopole-like singularity in a regular Hamiltonian is interesting but mysterious, and the implications of the resolution of the monopole singularity in the non-adiabatic domain have been recently discussed~\cite{Deguchi-Fujikawa}. It will be interesting to know the more details of the topology and topology change of Berry's phase.  We discuss this issue using an exactly solvable model~\cite{exact solution} which is defined by suitably choosing the parameters in the original model of Berry~\cite{Berry}.  A salient feature of Berry's connection ${\cal A}_{k}$  is
that the geometrical center of the monopole-like configuration and the origin of the Dirac string, which appears  when the net outgoing flux is nonvanishing, are displaced in the parameter space.  The magnetic charge of Berry's phase in the adiabatic domain is also quantized to be $e_{M}=2\pi\hbar$ consistent with the Wu-Yang gauge invariance condition. We discuss the smooth topology change from a monopole-like configuration to a dipole configuration, or rather the other way around, from  a dipole configuration to a monopole-like configuration, by combining this displacement and the quantized magnetic charge with Gauss' theorem
$\int_{S}(\nabla\times {\cal A})\cdot d\vec{S}=\int_{V} \nabla\cdot (\nabla\times {\cal A}) dV=0$ for a volume $V$ which is free of singularities. Gauss' theorem indicates that the monopole-like configuration  combined with the Dirac string is effectively a dipole. The smooth topology change from a dipole to a monopole then takes place when one regards the Dirac string associated with Berry's phase as unobservable if the Dirac string  satisfies the Wu-Yang gauge invariance condition. In the transitional region from a dipole to a monopole, a half-monopole with a magnetic charge $e_{M}/2$ appears and the Dirac string becomes observable analogously to the measurement of the Aharonov-Bohm phase of an electron using the magnetic flux generated by a superconducting current of the Cooper pair~\cite{Tonomura}.  

Some parameters are fixed to be time-independent in this solvable model associated with the original Berry's model~\cite{Berry}, but the effect of fixing these parameters turns out to be small in the present analysis of topology and topology change. This is explicitly illustrated by an  analysis of a generic model of Berry's phase. This property is consistent with the expectation that topological properties are not very sensitive to the smooth deformation of parameters. To our knowledge, no explicit analysis of the smooth topology change of Berry's phase, from a monopole to a dipole, has been given  in the past and our analysis will clarify the topological aspects of the monopole-like object in Berry's phase.

\section{Topology change in exactly solvable model}
We consider a magnetic moment placed in a rotating magnetic field $\vec{B}(t)$ which is the original model analyzed by Berry~\cite{Berry},
 but we choose a specific $\vec{B}(t)$ parameterized by $\varphi(t)=\omega t$ with constant $\omega$, and constant $B$ and $\theta$ with $\vec{\sigma}$ standing for Pauli matrices:
 \begin{eqnarray}\label{Hamiltonian}
&&\hat{H}=-\mu\hbar\vec{B}(t)\cdot\vec{\sigma},\nonumber\\
&&\vec{B}(t)=B(\sin\theta\cos\varphi(t), 
\sin\theta\sin\varphi(t),\cos\theta ).
\end{eqnarray}
The exact solution of the Schr\"{o}dinger equation
\begin{eqnarray}
i\hbar\partial_{t}\psi(t)=\hat{H}\psi(t)
\end{eqnarray}
 is then written as~\cite{exact solution},
\begin{eqnarray}\label{eq-exactamplitude1}
\psi_{\pm}(t)
&=&w_{\pm}(t)\exp\left[-\frac{i}{\hbar}\int_{0}^{t}dt
w_{\pm}^{\dagger}(t)\big(\hat{H}
-i\hbar\partial_{t}\big)w_{\pm}(t)\right]\nonumber\\
&=&w_{\pm}(t)\exp\left[-\frac{i}{\hbar}\int_{0}^{t}dt
w_{\pm}^{\dagger}(t)\hat{H}w_{\pm}(t)\right]\exp\left[-\frac{i}{\hbar}\int_{0}^{t}
\vec{{\cal A}}_{\pm}(\vec{B})\cdot\frac{d\vec{B}}{d t}dt\right]
\end{eqnarray}
where 
\begin{eqnarray}\label{exact eigenfuntion}
w_{+}(t)&=&\left(\begin{array}{c}
            \cos\frac{1}{2}(\theta-\alpha) e^{-i\varphi(t)}\\
            \sin\frac{1}{2}(\theta-\alpha)
            \end{array}\right), \ \ \ 
w_{-}(t)=\left(\begin{array}{c}
            \sin\frac{1}{2}(\theta-\alpha) e^{-i\varphi(t)}\\
            -\cos\frac{1}{2}(\theta-\alpha)
            \end{array}\right)
\end{eqnarray}
and $\vec{{\cal A}}_{\pm}(\vec{B})\equiv w_{\pm}^{\dagger}(t)(-i\hbar\frac{\partial}{\partial \vec{B}})w_{\pm}(t)$.
The parameter $\alpha(\theta)$ is defined by
\begin{eqnarray}\label{parameter}
\tan\alpha(\theta)=\frac{(\hbar\omega/2\mu\hbar B)\sin\theta}{1+(\hbar\omega/2\mu\hbar B)
\cos\theta} =\frac{\sin\theta}{\eta + \cos\theta}
\end{eqnarray}
with 
\begin{eqnarray}\label{parameter2}
\eta=\frac{2\mu\hbar B}{\hbar\omega}=\frac{\mu BT}{\pi}
\end{eqnarray}
when one defines the period $T=2\pi/\omega$.
It is important that $w_{\pm}(t)$, which define the exact solutions,  are different from the instantaneous eigenfunctions of the Hamiltonian $\hat{H}$ at time $t$ that are given by setting $\alpha=0$ in \eqref{exact eigenfuntion}.  This shows that the adiabatic approximation using the instantaneous eigenfunctions cannot describe the smooth topology change discussed below. 

The solution \eqref{eq-exactamplitude1} is confirmed by evaluating
\begin{eqnarray}
i\hbar\partial_{t}\psi_{\pm}(t)
&=&\{ i\hbar\partial_{t}w_{\pm}(t)+w_{\pm}(t)[w_{\pm}^{\dagger}(t)\big(\hat{H}
-i\hbar\partial_{t}\big)w_{\pm}(t)]\}\nonumber\\
&&\times\exp\left[-\frac{i}{\hbar}\int_{0}^{t}dt^{\prime}
w_{\pm}^{\dagger}(t^{\prime})\big(\hat{H}
-i\hbar\partial_{t^{\prime}}\big)w_{\pm}(t^{\prime})\right]\nonumber\\
&=&\{ i\hbar\partial_{t}w_{\pm}(t)+w_{\pm}(t)[w_{\pm}^{\dagger}(t)\big(\hat{H}
-i\hbar\partial_{t}\big)w_{\pm}(t)]\nonumber\\
&&\   +w_{\mp}(t)[w_{\mp}^{\dagger}(t)\big(\hat{H}
-i\hbar\partial_{t}\big)w_{\pm}(t)]\}\nonumber\\
&&\times\exp\left[-\frac{i}{\hbar}\int_{0}^{t}dt^{\prime}
w_{\pm}^{\dagger}(t^{\prime})\big(\hat{H}
-i\hbar\partial_{t^{\prime}}\big)w_{\pm}(t^{\prime})\right]\nonumber\\
&=&\hat{H}\psi_{\pm}(t)
\end{eqnarray}  
where we used, by noting \eqref{parameter},
\begin{eqnarray}   
w_{\mp}^{\dagger}\big(\hat{H}-i\hbar\partial_{t}\big)w_{\pm}=0
\end{eqnarray}
 and the completeness relation $w_{+}w_{+}^{\dagger}+w_{-}w_{-}^{\dagger}=1$.

The quantity in \eqref{eq-exactamplitude1}
\begin{eqnarray}\label{connection}
\vec{{\cal A}}_{\pm}(\vec{B})\equiv w_{\pm}^{\dagger}(t)(-i\hbar\frac{\partial}{\partial \vec{B}})w_{\pm}(t)
\end{eqnarray}
gives an analogue of the gauge potential (or connection) in the parameter space. 
The extra phase factor for one period of motion is given by,
\begin{eqnarray}\label{Berry's phase1}
\exp\left[-\frac{i}{\hbar}\oint
\vec{{\cal A}}_{\pm}(\vec{B})\cdot\frac{d\vec{B}}{d t}dt\right]&=&\exp\{-i\oint \frac{-1\mp\cos(\theta-\alpha(\theta))}{2}d\varphi \}\nonumber\\
&=&\exp\{-i\oint\frac{1\mp\cos(\theta-\alpha(\theta))}{2}d\varphi+2i\pi \}\nonumber\\
&=&\exp\{-\frac{i}{\hbar}\Omega_{\pm} \},
\end{eqnarray}
with the monopole-like flux
\begin{eqnarray}\label{solid-angle}
\Omega_{\pm} &=& \hbar\oint  \frac{(1\mp\cos(\theta-\alpha(\theta)))}{2}d\varphi\nonumber\\
&=& 2\pi\hbar \frac{(1\mp\cos(\theta-\alpha(\theta)))}{2}.
\end{eqnarray}
In \eqref{Berry's phase1}, we adjusted the trivial phase $2\pi i$ for the convenience of the later analysis; this is related to a  gauge transformation of Wu and Yang~\cite{Wu-Yang} discussed below.   From now on, we concentrate on $\Omega_{+}$.

\subsection{Classification of topological configurations}

From \eqref{solid-angle}, we have the monopole-like potential
\begin{eqnarray}\label{potential1}
{\cal A}_{\varphi} 
= \frac{\hbar }{2B\sin\theta} \left(1 - \cos(\theta-\alpha(\theta)) \right)
\end{eqnarray}
and ${\cal A}_{\theta} ={\cal A}_{B} =0$. We have $\hbar$ in \eqref{potential1} which shows that the potential is an order $O(\hbar)$ quantum effect in the present context. We want to clarify precisely what kind of object is described by the potential \eqref{potential1}.  

We start with the analysis of the parameter $\alpha(\theta)$.
In Fig.1, we show the relation between $\theta$ and $\tan\alpha(\theta)$ for the case $0\leq \eta<1$ given by \eqref{parameter}.  
\begin{figure}[htb]
 \begin{center}
   \includegraphics*[width = 7.cm]{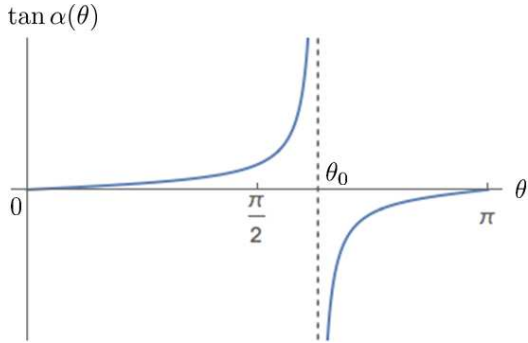}
   \caption{The relation between $\theta$
and $\tan\alpha(\theta)$ determined by Eq. \eqref{parameter} for $0\leq \eta<1$ with $\cos\theta_{0}=-\eta$.}\label{Fig_01}
 \end{center}
\end{figure}
For this parameter range, we have a singularity at $\cos\theta_{0}=-\eta$ in the denominator of \eqref{parameter}. But this does not give rise to a singular relation between $\alpha(\theta)$ and $\theta$; one can confirm
\begin{eqnarray}\label{slope}
\frac{d\alpha(\theta)}{d\theta}=\frac{1+\eta\cos\theta}{(\eta+\cos\theta)^{2}+\sin^{2}\theta}
\end{eqnarray}
and thus 
\begin{eqnarray}\label{smoothness}
\frac{d\alpha(\theta)}{d\theta}|_{\theta=\theta_{0}}=1
\end{eqnarray} 
for $\cos\theta_{0}=-\eta$.
For the parameter range $\eta\geq 1$,  the relation \eqref{parameter} is smooth. For $\eta=1$, we have an exact relation
\begin{eqnarray}
\alpha(\theta)=\theta/2.
\end{eqnarray}
 For other parameter values, we have 
\begin{eqnarray}\label{limiting form}
&&\alpha(\theta)=\frac{1}{\eta}\sin\theta  \ \ \ \ {\rm for} \ \eta \gg 1,\nonumber\\
&&\alpha(\theta)=\theta -\eta \sin\theta  \ \ \ \ {\rm for} \ 0\leq \eta \ll 1.
\end{eqnarray}

In the following analysis of topology, it will be shown that the value of $\eta=\mu BT/\pi$ in \eqref{parameter2} plays a central role to specify topology, namely, invariance under the smooth variation of parameters.  The parameter domain $\eta>1$ defines the adiabatic domain and implies the existence of a monopole-like configuration regardless of the values of $B$ and $T$ individually; ``adiabatic'' implies typically large $T$ with fixed $B$. The domain $0\leq \eta<1$ defines the non-adiabatic domain  and implies the appearance of a dipole-like configuration (and the disappearance of a monopole-like configuration) regardless of the values of $B$ and $T$ individually; ``non-adiabatic'' implies typically small $T$ with fixed $B$.  

In the analysis of topology change, the transition from $\eta>1$ to
$\eta<1$ through the critical value $\eta=1$ is important. In Fig.2, we thus show the relation between $\alpha(\theta)$ and $\theta$ at the transition region near $\eta=1$ given by \eqref{parameter}. 
\begin{figure}[htb]
 \begin{center}
   \includegraphics*[width = 8.0cm]{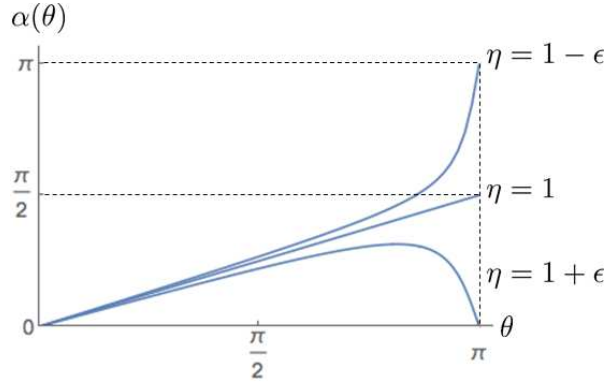}
   \caption{The topology change at the parameter value $\eta=1$ determined by Eq. \eqref{parameter}.}\label{Fig_02}
 \end{center}
\end{figure}
For the parameters
$\eta=1\pm\epsilon$
with a small positive $\epsilon$, the value $\alpha(\theta)$ departs from the common value $\frac{1}{2}\theta$ assumed at around $\theta=0$ and splits into two branches for the values of the parameter $\theta$ close to  $\theta=\pi$. We have $\alpha(\pi)=0$ for $\eta=1+\epsilon$ and  $\alpha(\pi)=\pi$ for $\eta=1-\epsilon$, respectively,  with the slopes
\begin{eqnarray}\label{singularity}
\frac{d\alpha(\theta)}{d\theta}|_{\theta=\pi} =\mp \frac{1}{\epsilon}
\end{eqnarray}
for $\eta=1\pm \epsilon$, respectively, using \eqref{slope}. We thus observe the singular jump characteristic to the topology change in terms of $\alpha(\theta)$ at $\eta=1$. 

When one defines
\begin{eqnarray}
\Theta(\theta; \eta)=\theta-\alpha(\theta),
\end{eqnarray}
but without writing $\eta$ explicitly, we have $\Theta(0)=0$  and 
\begin{eqnarray}\label{end values}
\Theta(\pi)=\pi, \ \pi/2, \ 0
\end{eqnarray}
respectively, for $\eta>1$, $\eta=1$, and $\eta<1$.  We also have 
\begin{eqnarray}\label{turning point1}
\frac{\partial\Theta(\theta)}{\partial\theta}|_{\theta=\theta_{0}}=0
\end{eqnarray}
for $\eta<1$ using \eqref{smoothness}. 
In Fig.3, we show the relation between $\theta$ and $\Theta(\theta)$.
\begin{figure}[htb]
 \begin{center}
   \includegraphics*[width = 8.0cm]{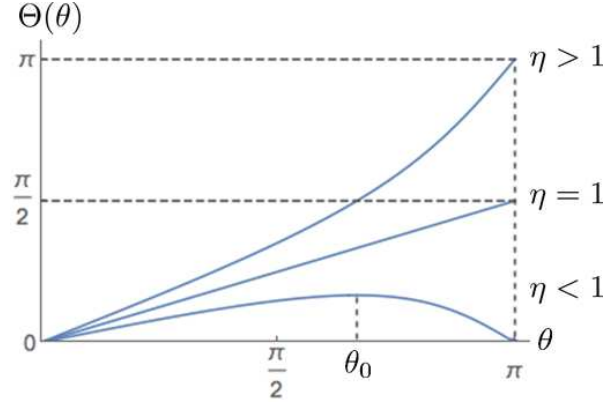}
   \caption{The relation between $\theta$
and $\Theta(\theta)$ parameterized by $\eta$. Note that $\cos\theta_{0}=-\eta$.}\label{Fig_03}
 \end{center}
\end{figure}
We write the monopole-like potential \eqref{potential1} in the form  
\begin{eqnarray}\label{exact potential}
{\cal A}_{\varphi} 
=  \frac{\hbar}{2B\sin\theta}(1 - \cos\Theta(\theta)).
\end{eqnarray}
The variable $\Theta(\theta)$ thus describes the essence of the topology and topology change of Berry's phase.
The topology change is seen in the change of $\Theta(\pi)=\pi$ for $\eta>1$ to $\Theta(\pi)=0$ for $\eta<1$ in Fig.3.  But we have a well-defined potential at the boundary $\eta=1$
\begin{eqnarray}\label{half sphere}
{\cal A}_{\varphi}&=&\frac{\hbar}{2B\sin\theta}(1-\cos\frac{1}{2}\theta)
\end{eqnarray}  
for $\theta\neq \pi$. We also note that the Dirac string which corresponds to the singularity of the potential \eqref{exact potential} can appear at $\theta=0$ or $\theta=\pi$; no singularity at $\theta=0$ since $\Theta(0)=0$, and the possible Dirac string appears at $\theta=\pi$ for $\Theta(\pi)=\pi$ ($\eta>1$) or $\Theta(\pi)=\pi/2$ ($\eta=1$) but no string for $\Theta(\pi)=0$ ($\eta<1$).

Using the exact potential \eqref{exact potential} and ${\cal A}_{\theta}={\cal A}_{B}=0$,
we have an analogue of the magnetic flux in the parameter space $\vec{B}=B(\sin\theta\cos\varphi, 
\sin\theta\sin\varphi,\cos\theta )$,
\begin{eqnarray}\label{effective magnetic flux}
\nabla\times {\cal A}|_{\eta}
&=&\frac{\hbar}{2}\frac{\Theta^{\prime}(\theta)\sin\Theta(\theta)}{\sin\theta}\frac{1}{B^{2}}{\bf e}_{B}
\end{eqnarray}
  for $\theta\neq \pi$ with $\Theta^{\prime}(\theta)=\frac{\partial\Theta(\theta)}{\partial\theta}$.  In this evaluation of the flux, we keep the parameter $\eta=\mu TB/\pi$ fixed, since $\Theta(\theta)=\Theta(\theta;\eta)$. 
It is significant that the ``magnetic flux'' is always pointing in the radial direction ${\bf e}_{B}=\frac{\vec{B}}{B}$, but the magnitude of the flux depends on the angle $\theta$.  

As for the integrated net outgoing  flux from a sphere centered at $\vec{B}=0$, 
avoiding the singular point $\theta=\pi$, we have
\begin{eqnarray}\label{total flux}
\int_{\theta\neq\pi}\nabla\times {\cal A}|_{\eta}\cdot d\vec{S}&=&\int \frac{\hbar}{2}\frac{\Theta^{\prime}(\theta)\sin\Theta(\theta)}{\sin\theta}\frac{1}{B^{2}}B^{2}\sin\theta d\varphi d\theta\nonumber\\
&=&\int^{\pi}_{0} \frac{2\pi\hbar}{2}\Theta^{\prime}(\theta)\sin\Theta(\theta) d\theta \nonumber\\
&=&\pi\hbar (1-\cos\Theta(\pi))
\end{eqnarray}
which agrees with Stokes' theorem applied to \eqref{exact potential} near the south pole.
  
We now illustrate the typical topological configurations from the point of view 
of the outgoing flux. In the adiabatic limit 
$\eta=\mu BT/\pi\rightarrow\infty$ (i.e., $T\rightarrow \infty$ with fixed $B$), we have $\Theta(\theta)\rightarrow \theta$ due to \eqref{limiting form}, and we have the Dirac monopole-like flux
\begin{eqnarray}\label{adiabatic flux}
\nabla\times {\cal A}|_{\eta}
&=&\frac{e_{M}}{4\pi}\frac{1}{B^{2}}{\bf e}_{B}
\end{eqnarray}  
with the magnetic charge $e_{M}=2\pi\hbar$.
We thus have the integrated flux
\begin{eqnarray}
\int_{\theta\neq\pi}\nabla\times {\cal A}|_{\eta}\cdot d\vec{S}&=&
e_{M}.
\end{eqnarray}
In the transitional domain 
$\eta=\mu BT/\pi=1$, we have $\Theta(\theta)=\frac{1}{2} \theta$, and we have the flux
\begin{eqnarray}\label{transitional flux}
\nabla\times {\cal A}|_{\eta}
&=&\frac{e_{M}}{8\pi}\frac{\sin\frac{1}{2}\theta}{\sin\theta}\frac{1}{B^{2}}{\bf e}_{B}
\end{eqnarray}   
which is pointing to the direction of ${\bf e}_{B}=\frac{\vec{B}}{B}$ but its magnitude depends on the angle $\theta$ and divergent for $\theta\rightarrow \pi$.
The integrated  flux is, however, finite and half of the value of the adiabatic limit 
\begin{eqnarray}\label{half-monopole}
\int_{\theta\neq\pi}\nabla\times {\cal A}|_{\eta}\cdot d\vec{S}&=&
\frac{1}{2}e_{M}.
\end{eqnarray}
In the non-adiabatic limit, $\eta=\mu BT/\pi\rightarrow0$ (i.e., $T\rightarrow 0$ with fixed $B$), we have $\Theta(\theta)\rightarrow 0$ due to \eqref{limiting form}, and thus 
 \begin{eqnarray}\label{non-adiabatic flux}
\nabla\times {\cal A}|_{\eta}
&\rightarrow&0,
\end{eqnarray}
namely, the monopole-like object disappears.
We thus recognize three distinct topological configurations.  

To visualize the topology specified by the value of $\eta$, we draw  schematic figures in Fig.4a $\sim$ 4c which are based on the formula \eqref{exact potential} with the integrated flux \eqref{total flux} and the movement of $\Theta(\theta)$ in Fig.3.
\begin{figure}[htb]
 \begin{center}
   \includegraphics*[width = 14.cm]{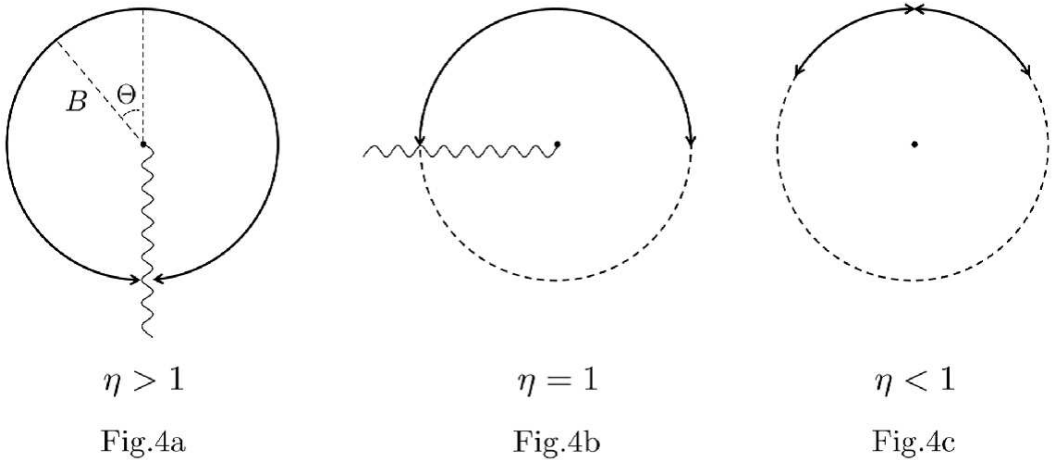}
   \caption{Fixed $\eta=\mu BT/\pi$ pictures. The wavy lines symbolically represent the Dirac strings.}\label{Fig_04a_b_c}
 \end{center}
\end{figure}
When one varies the parameter $\theta$ from $0$ to $\pi$, one has a full coverage of the sphere $S^{2}$ with the appearance of a Dirac string at $\theta=\pi$ for $\eta>1$ (adiabatic domain) in Fig.4a since $\Theta(\pi)=\pi$, which is analogous to the Dirac monopole.  The wavy line in Fig.4a  along the negative $z$-axis represents the Dirac string.  Note that fixed $\eta=\mu BT/\pi$ means that $T$ varies when one changes $B$, in contrast to the fixed $T$ figures in Fig.5 and Fig.6a $\sim$ 6b discussed later.

For $\eta=1$ (transitional domain) in Fig.4b, we have a half covering of $S^{2}$ but still with a Dirac string at $\theta=\pi$ since $\Theta(\pi)=\pi/2$. We show schematically the Dirac string by a wavy line along the negative $x$-axis  in Fig.4b. 
For $\eta<1$ (non-adiabatic domain) in Fig.4c, we have no covering and no Dirac string since $\Theta(0)=\Theta(\pi)=0$ (the turning point of the arrows in the figure takes place at $\theta=\theta_{0}$ in \eqref{turning point1}), which is topologically identified to be a dipole as will be explained in more detail later. 

The new ingredient in the present  analysis, which was absent in the analysis of $\Theta(\theta)$ in Fig.3, is the appearance of the Dirac string at $\theta=\pi$.
It is important that both singular behaviors \eqref{singularity} and the Dirac string  appear at $\theta=\pi$. In other words, one can choose Berry's phase to be regular for $\theta\neq\pi$.

In passing, we mention that when one varies $T$ for fixed B, one observes the configurations in Fig.4 starting with Fig.4a to Fig.4b and then to Fig.4c, corresponding to the change of $T$ from $T\rightarrow \infty$ ($\eta=\infty$) to $T=\pi/\mu B$ ($\eta=1$) and then to $T\rightarrow 0$ ($\eta=0$), respectively.

\subsection{Smooth topology change}

We have useful information about the topology change from  Gauss' theorem which states that 
\begin{eqnarray}\label{Gauss}
\int_{S}(\nabla\times {\cal A})\cdot d\vec{S}=\int_{V} \nabla\cdot (\nabla\times {\cal A}) dV=0
\end{eqnarray}
using the formula of vector analysis
\begin{eqnarray}
\nabla\cdot (\nabla\times {\cal A}) =0.
\end{eqnarray}
Here the volume $V$ is defined by excluding a thin tube covering the Dirac string as in Fig.5 and $S$ stands for the surface of this volume $V$ for a fixed value of $T$.
\begin{figure}[htb]
 \begin{center}
   \includegraphics*[width = 6.0cm]{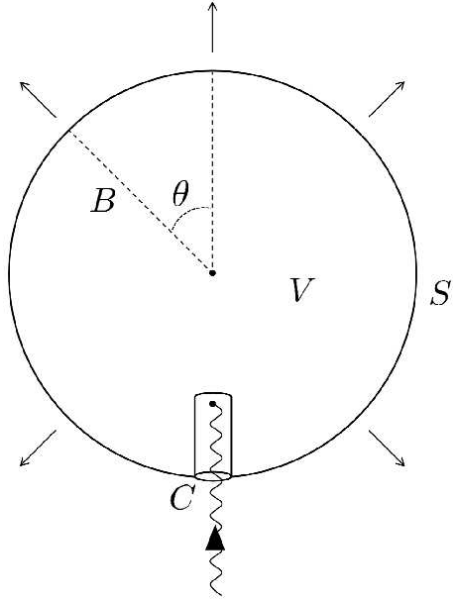}
   \caption{Fixed $T$ picture.  Volume V avoids a thin tube surrounding the Dirac string. Geometrical center and the origin of the Dirac string are displaced by the distance $B=\pi/\mu T$.}\label{Fig_05}
 \end{center}
\end{figure}
Note that there is no singularity inside the volume $V$.
The Dirac string originates at $z=-\pi/\mu T$
on the negative z-axis
corresponding to $\eta=\mu TB/\pi=1$ with fixed $T$. Recall that no Dirac string appears for $\eta<1$
since $\Theta(0)=\Theta(\pi)=0$ as in Fig.3 and thus no singularity in \eqref{exact potential} at $\theta=0$ or $\theta=\pi$. The present fixed $T$ picture is convenient to understand the difference between Berry's phase and the genuine Dirac monopole.  

For the fixed $T$ picture, we have instead of \eqref{effective magnetic flux}
\begin{eqnarray}\label{magnetic flux}
\nabla\times {\cal A}
&=&\frac{\hbar}{2}\frac{\Theta^{\prime}(\theta)\sin\Theta(\theta)}{\sin\theta}\frac{1}{B^{2}}{\bf e}_{B} - \frac{\hbar}{2}\frac{\frac{\partial\Theta(\theta)}{\partial B}\sin\Theta(\theta)}{B\sin\theta}{\bf e}_{\theta}
\end{eqnarray}
where ${\bf e}_{\theta}$ is a unit vector in the direction $\theta$ in the spherical coordinates.
By recalling $\Theta(\theta)=\Theta(\theta;\eta)$  with $\eta=\mu TB/\pi$, we have
\begin{eqnarray}
\frac{\partial\Theta(\theta)}{\partial B}=\frac{\partial\eta}{\partial B}
\frac{\partial\Theta(\theta)}{\partial \eta}=\frac{\mu T}{\pi}\frac{\partial\Theta(\theta)}{\partial \eta}
\end{eqnarray}
and using $\Theta(\theta)=\theta-\alpha(\theta)$ and \eqref{parameter}
\begin{eqnarray}
\frac{\partial\Theta(\theta)}{\partial \eta}=\frac{\sin\theta}{(\eta+\cos\theta)^{2}+\sin^{2}\theta}.
\end{eqnarray}

The discrepancy of \eqref{total flux} and \eqref{Gauss} is attributed to the contribution 
of the Dirac string.  
It is useful to confirm Gauss' theorem in the present context for the adiabatic domain $\eta>1$.  The first term in \eqref{magnetic flux}
determines the contribution from the outer surface in Fig.5 
\begin{eqnarray}\label{outer surface}
\int_{S_{out}}(\nabla\times {\cal A})\cdot d\vec{S}=2\pi\hbar
\end{eqnarray}
using the result in \eqref{total flux}.
The second term in \eqref{magnetic flux} describes a contribution of the cylinder part of the thin tube surrounding the Dirac string in Fig.5
\begin{eqnarray}\label{cylinder}
\int (\nabla\times {\cal A})\cdot dS_{\theta}
&=& - \int \frac{\hbar}{2}\frac{\frac{\partial\Theta(\theta)}{\partial B}\sin\Theta(\theta)}{B\sin\theta}dB B\sin\theta d\varphi\nonumber\\
&=&-\pi\hbar\int_{\pi/\mu T}^{B}\frac{\partial\Theta(\theta)}{\partial B}\sin\Theta(\theta)dB\nonumber\\
&=&\pi\hbar(\cos\Theta(\theta;\eta)-\cos\Theta(\theta;\eta=1))\nonumber\\
&=&\pi\hbar(\cos\Theta(\theta)-\cos\frac{1}{2}\theta)
\end{eqnarray}
using  the surface element
\begin{eqnarray}
dS_{\theta}=dB B\sin\theta d\varphi {\bf e}_{\theta}
\end{eqnarray}
and $\Theta(\theta;\eta=1)=\frac{1}{2}\theta$.
As for the contribution of a small cap around the origin of the Dirac string in Fig.5, we ``blow it up'' to a full surface without encountering a singularity.  The picture is then analogous to the outer surface in Fig.6b discussed below, but the  inside of the sphere is outside  the volume $V$, and thus the contribution from the blown-up sphere is given by 
\begin{eqnarray}\label{cap}
-\pi\hbar (1-\cos\frac{1}{2}\theta)
\end{eqnarray}
from \eqref{total flux} (but with a free value of $\theta$ without fixing it at $\theta=\pi$ for the moment) using $\Theta(\theta)=\frac{1}{2}\theta$ for $\eta=1$.  The sum of \eqref{cylinder} and \eqref{cap} gives
\begin{eqnarray}
\pi\hbar(\cos\Theta(\theta)-\cos\frac{1}{2}\theta)-\pi\hbar (1-\cos\frac{1}{2}\theta)
=-\pi\hbar(1-\cos\Theta(\theta))
\end{eqnarray}
which gives $-2\pi\hbar$ when one sets $\theta=\pi$ and cancels the contribution from the outer surface \eqref{outer surface} in Fig.5, in agreement with Gauss' theorem \eqref{Gauss}.

More formally, Stokes' theorem states in the adiabatic domain $\eta>1$ using \eqref{total flux}
\begin{eqnarray}\label{Stokes1}
\oint_{C} {\cal A}_{\varphi}B\sin\theta d\varphi =\int_{S^{\prime}}(\nabla\times {\cal A})\cdot d\vec{S}=2\pi\hbar
\end{eqnarray}
for an infinitesimally small circle $C$ surrounding the Dirac string in Fig.5. This flux is regarded, depending on the choice of $S^{\prime}$, either as the flux flowing out of the volume $V$ indicated by \eqref{total flux} or the flux flowing into the volume $V$ through the Dirac string by recalling the fact that no singularity exists inside the volume $V$ in Fig.5.   
      
The surface $S$ on the left-hand side of Gauss' theorem \eqref{Gauss} does not cover the singularity and in this sense topologically trivial.  The Gauss' theorem \eqref{Gauss} is valid for a smooth decrease of $B$ starting with Fig.6a to Fig.6b and then to Fig.6c. 
\begin{figure}[htb]
 \begin{center}
   \includegraphics*[width = 14.cm]{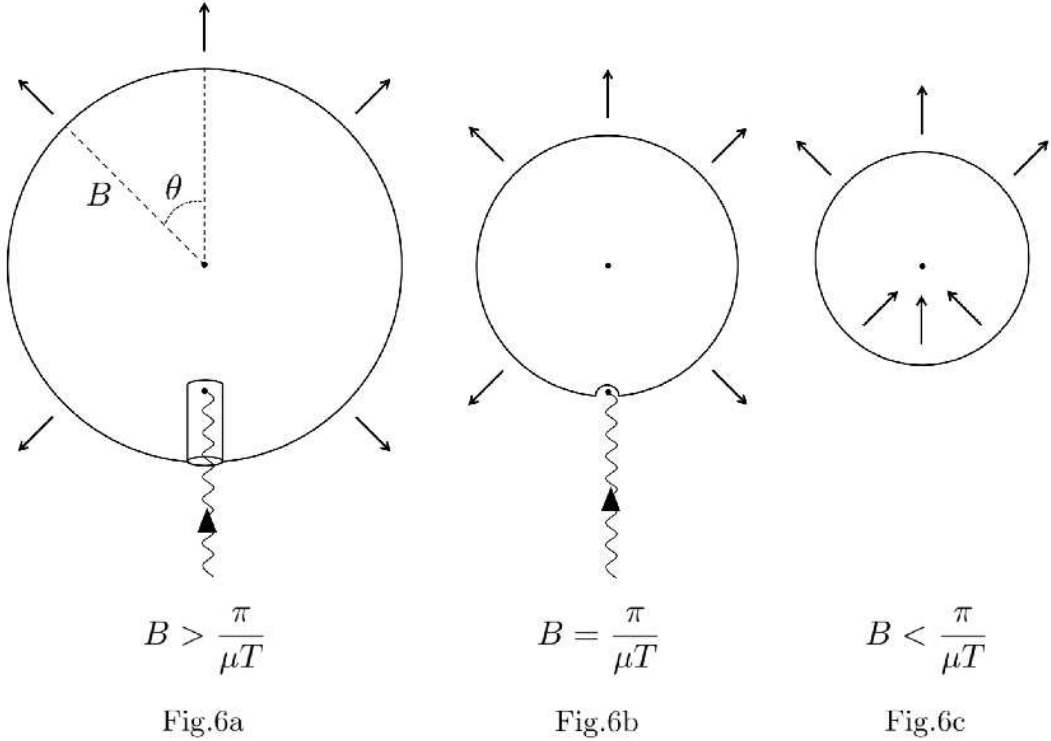}
   \caption{Fixed $T$ pictures with varying radius $B$ (and thus varying $\eta=\mu TB/\pi$).  Geometrical center and the origin of the Dirac string are displaced by the distance $B=\pi/\mu T$.}\label{Fig_06a_b_c}
 \end{center}
\end{figure}
We shall argue that the origin of the smooth topology change in Berry's phase resides in this trivial topology for all the topological configurations.
Using  the second expression in \eqref{total flux}
\begin{eqnarray}\label{differential flux}
\frac{2\pi\hbar}{2}\Theta^{\prime}(\theta)\sin\Theta(\theta) 
\end{eqnarray}
and the movement of $\Theta(\theta)$ in Fig.3, 
we show the schematic  pictures with fixed $T$ in Fig.6a$\sim$ 6c. We here use the parameter $\theta$ which covers the full range from $0$ to $\pi$ for all the cases in Fig.6a$\sim$ 6c. 
Both the Dirac monopole-like flux and the Dirac string indicated by a wavy line are seen when observed at $B>\pi/(\mu T)$ (adiabatic domain) in Fig.6a. One has the transitional                                                                                                                                 domain at $B=\pi/(\mu T)$ in Fig.6b where both the out-going flux from an outer sphere and a small half-sphere covering the origin of the Dirac string are still seen, although half of the strength of those in Fig.6a. See \eqref{half-monopole}. No net outgoing flux and no Dirac string are observed when one comes closer to the monopole position $B<\pi/(\mu T)$ (non-adiabatic domain) in Fig.6c, which looks like the flux from a small Earth (i.e., a dipole).  The inward flux in Fig.6c arises from the negative signature of 
\begin{eqnarray}
\Theta^{\prime}(\theta)=\frac{\partial\Theta(\theta)}{\partial\theta}<0
\end{eqnarray}
in \eqref{differential flux} for $\eta<1$ and $\theta_{0}<\theta$. See Fig.3.  It is remarkable that the topological properties of the monopole-like object in Berry's phase are very rich.

From a point of view of the net outgoing flux, we thus see the full flux with $e_{M}=2\pi\hbar$ in Fig.6a and the half flux with $e_{M}/2$ in Fig.6b and then no net flux in Fig.6c, corresponding to $\Theta(\pi)$ with $\pi$, $\pi/2$ and $0$, respectively, in \eqref{total flux}. 
Thus these configurations are very distinct. 

On the other hand, Gauss' theorem \eqref{Gauss} shows a smooth transition among distinct topologies specified by $\Theta(\pi)$ with $\pi$, $\pi/2$ and $0$.  Our smoothness argument of topology change in Berry's phase is based on the Gauss theorem but we use the arguments of Dirac \cite{Dirac} and Wu and Yang\cite{Wu-Yang} to distinguish different configurations. Namely, if the Dirac string is not observable, then we ignore it physically and identify a monopole.  This unobservability critically depends on the magnetic charge of the monopole-like object and leads to the quantization of the charge in the case of the genuine Dirac monopole~\cite{Dirac, Wu-Yang}. In the present case, the magnetic charge is fixed by the formula of Berry's phase.  Thus if the magnetic flux carried by the Dirac string satisfies the unobservability condition, we regard the monopole-like object as a physical monopole, and otherwise no physical monopole, namely, we regard a combination of the monopole-like object accompanied by the string as a physical entity.

We start with an analysis of the adiabatic configuration with $\eta=\mu TB/\pi >1$ such as in Fig.6a.  The argument of Wu and Yang is to consider the singularity-free potentials in the upper and lower hemispheres        
\begin{eqnarray}
{\cal A}_{\varphi +} 
&=&  \frac{e_{M}}{4\pi B\sin\theta}(1 - \cos\Theta(\theta)),\nonumber\\
{\cal A}_{\varphi -} 
&=&  \frac{e_{M}}{4\pi B\sin\theta}(- 1 - \cos\Theta(\theta)),
\end{eqnarray}
using the potential in \eqref{exact potential} with $e_{M}=2\pi\hbar$. 
These two potentials are related by a gauge transformation 
\begin{eqnarray}
{\cal A}_{\varphi -}={\cal A}_{\varphi +} - \frac{\partial\Lambda}{B\sin\theta\partial \varphi}
\end{eqnarray}
with 
\begin{eqnarray}
\Lambda=\frac{e_{M}}{2\pi} \varphi.
\end{eqnarray}
The physical condition is 
\begin{eqnarray}
\exp[-\frac{i}{\hbar}\oint {\cal A}_{\varphi -}B\sin\theta d\varphi]&=&\exp[-\frac{i}{\hbar}\oint{\cal A}_{\varphi +}B\sin\theta d\varphi +\frac{i}{\hbar}\oint \frac{\partial\Lambda}{B\sin\theta\partial \varphi}B\sin\theta d\varphi]\nonumber\\
&=&\exp[-\frac{i}{\hbar}\oint{\cal A}_{\varphi +}B\sin\theta d\varphi]
\end{eqnarray}
which is in fact satisfied since the gauge term gives
$\exp[i e_{M}/\hbar]=\exp[2\pi i]=1$ and thus defines a monopole. 
Note that the physical condition in the present context is that the Schr\"{o}dinger wave function \eqref{eq-exactamplitude1} is single valued under the gauge transformation. It is confirmed that the present argument of gauge transformation is equivalent to the evaluation of the phase change induced by the Dirac string~\cite{Wu-Yang}. 
The fact that the physical condition is satisfied 
shows that the magnetic charge 
\begin{eqnarray}\label{magnetic charge}
e_{M}=2\pi \hbar
\end{eqnarray}
is properly quantized satisfying the Dirac quantization condition, although we have no analogue of an electric coupling in the present case unlike the original Dirac monopole~\cite{Dirac}.

In contrast, for the transitional domain $\eta=\mu TB/\pi =1$
such as in Fig.6b 
we have two potentials from \eqref{half sphere}  
\begin{eqnarray}
{\cal A}_{\varphi +}&=&\frac{e_{M}}{4B\sin\theta}(1-\cos\frac{1}{2}\theta)\nonumber\\
{\cal A}_{\varphi -}&=&\frac{e_{M}}{4B\sin\theta}(-\cos\frac{1}{2}\theta)
\end{eqnarray}
which are well-defined in the upper and lower hemispheres, respectively, and are related by the gauge transformation 
\begin{eqnarray}
{\cal A}_{\varphi -}={\cal A}_{\varphi +} - \frac{\partial\Lambda}{B\sin\theta\partial \varphi}
\end{eqnarray}
with 
\begin{eqnarray}
\Lambda=\frac{e_{M}}{4\pi} \varphi.
\end{eqnarray}
The physical condition 
\begin{eqnarray}
\exp[-\frac{i}{\hbar}\oint {\cal A}_{\varphi -}B\sin\theta d\varphi]&=&\exp[-\frac{i}{\hbar}\oint{\cal A}_{\varphi +}B\sin\theta d\varphi +\frac{i}{\hbar}\oint \frac{\partial\Lambda}{B\sin\theta\partial \varphi}B\sin\theta d\varphi]\nonumber\\
&=&\exp[-\frac{i}{\hbar}\oint{\cal A}_{\varphi +}B\sin\theta d\varphi]
\end{eqnarray}
is not satisfied since the gauge transformation gives
\begin{eqnarray}\label{half monopole phase}
\exp[ie_{M}/2\hbar]=\exp[i\pi ]=-1.
\end{eqnarray}
We thus conclude that the half-monopole at the transitional domain $\eta=1$ with the magnetic charge $e_{M}/2$ cannot describe a physical monopole; it is physical  as a combination of  the monopole-like object, which generates the outgoing flux, accompanied by a Dirac string  \footnote{A half-monopole with a magnetic charge $e_{M}/2$ gives a non-trivial phase \eqref{half monopole phase} and thus the Dirac string is not unobservable. In fact this phase of $\exp[i\pi]$ is the same as the Aharonov-Bohm phase of an electron in the magnetic field generated by the superconducting current of the Cooper pair in the experiment by Tonomura~\cite{Tonomura}.  In our criterion following the analysis of Wu and Yang \cite{Wu-Yang}, the Dirac string thus becomes a physical
observable just as the outgoing flux from the monopole-like object. }, although the Dirac string is actually defined only at $B=\pi/\mu T$. Topologically, it is thus the same as the dipole for $\eta<1$ in Fig.6c. 

In fact, from the point of view of the Gauss' theorem \eqref{Gauss}, all the configurations of Berry's phase are topologically the dipole as is seen in Fig.5; the monopole is identified only when the Dirac string satisfies the Wu-Yang gauge invariance condition, or equivalently Dirac's quantization condition, and thus becomes unobservable. This is a mechanism of the smooth topology change in Berry's phase when one approaches the monopole position in the parameter apace.

In comparison, we show a schematic figure of a genuine Dirac monopole located at the origin of the parameter space in Fig.7. 
\begin{figure}[htb]
 \begin{center}
   \includegraphics*[width = 6.0cm]{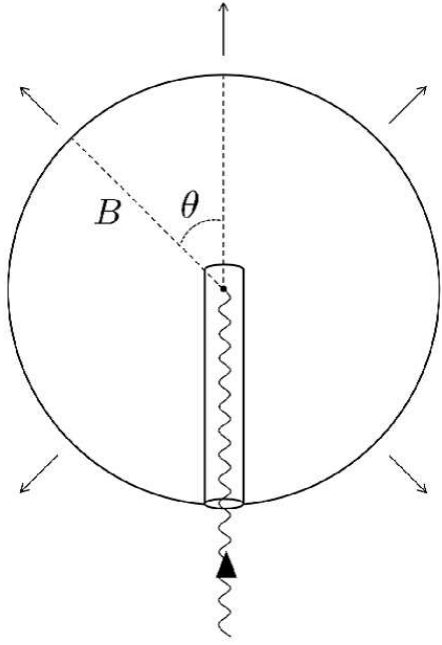}
   \caption{Genuine Dirac monopole in the parameter space.}\label{Fig_07}
 \end{center}
\end{figure}
For any fixed  value of $T$, we have the same figure as in Fig.7 for any value of $B$ with a Dirac string, which satisfies the Wu-Yang condition, stretching from the origin $\vec{B}=0$ of the parameter space to infinity. Thus no topology change from a monopole to a dipole takes place. From Fig.5 and Fig.7, one can see a clear difference between Berry's phase and a genuine Dirac monopole.

\subsection{Explicit forms of Berry's phase}

Finally we comment on the more explicit forms of Berry's phase which may be useful in practical applications. In  
the adiabatic limit $T=2\pi/\omega\rightarrow \infty$ (with $B\neq 0$),
\begin{eqnarray}
\eta=\frac{\mu TB}{\pi} \rightarrow \infty,
\end{eqnarray}
and the parameter $\alpha \rightarrow 0$ in \eqref{parameter}.  Berry's phase then gives
\begin{eqnarray}\label{Dirac monopole}
{\cal A}_{\varphi} 
=  \frac{\hbar}{2B\sin\theta}(1 - \cos\theta),
\end{eqnarray}
namely, one obtains the Dirac monopole-type potential in the parameter space $\vec{B}$. But it is important to recognize that the Dirac monopole-like configuration of Berry's phase  is realized only in the constrained one-dimensional sub-space $\mu TB/\pi=\infty$ ($T=\infty$ and finite $B$ in the present example~\cite{Simon}) unlike the full two-dimensional parameter space $(B,T)$ in the case of the genuine Dirac monopole.

In contrast, in the non-adiabatic limit $T=2\pi/\omega\rightarrow 0$ (or $B\rightarrow 0$),
\begin{eqnarray}
\eta = \frac{\mu TB}{\pi} \rightarrow 0,
\end{eqnarray}
and then the parameter $\Theta=\theta-\alpha \rightarrow 0$ using \eqref{limiting form}. The potential associated with Berry's phase \eqref{exact potential} thus becomes trivial
\begin{eqnarray}\label{trivial Berry phase}
{\cal A}_{\varphi} 
= 0.
\end{eqnarray}
Namely, Berry's phase is smoothly connected to a  trivial value for a continuous variation of $B\rightarrow 0$ with finite T, as is physically expected for the vanishing (real) magnetic field $B$ with  fixed $\omega$ in Berry's model \eqref{Hamiltonian}~\cite{Berry}.  To be more explicit, we have a useful relation in the non-adiabatic domain $\eta \ll 1$ using \eqref{limiting form},
\begin{eqnarray}
{\cal A}_{\varphi}&=&\frac{\hbar}{2B\sin\theta}(1-\cos(\theta-\alpha))\nonumber\\
&\simeq& \frac{\hbar}{4B}(\mu TB/\pi)^{2}\sin\theta
\end{eqnarray}  
that has no singularity associated with the Dirac string at $\theta=\pi$.

\section{Generic model}

 We have so far analyzed an exactly solvable model. We now discuss the generality of the results obtained by the  specific model.  The generic model of Berry's phase is given by the model~\footnote{Note that $|\vec{p}(t)|$ corresponds to $\hbar B$ in the exactly solvable model  \eqref{Hamiltonian}, and the parameter $\eta$ in \eqref{parameter2} is replaced by 
 $\eta=\mu T|\vec{p}(t)|/\pi\hbar$. The magnetic potential and flux are defined by $\oint \vec{{\cal A}}\cdot d\vec{p}$ to conform to the convention in \cite{Deguchi-Fujikawa}, in comparison to $\oint \vec{{\cal A}}\cdot d\vec{B}$ in \eqref{Hamiltonian}.} 
\begin{eqnarray}\label{generic Hamiltonian}
H=-\mu \vec{\sigma}\cdot\vec{p}(t)
\end{eqnarray} 
which appears in the context of band-crossing
problems in condensed matter physics~\cite{MacDonald, Hirsch}.  This model is also related to the model analyzed by Stone~\cite{Stone}, and also to the general level-crossing problem which have been analyzed using the technique of second quantization~\cite{fujikawa-mpla2005, Deguchi}.  In condensed matter physics $\vec{p}(t)$ stands for the Bloch momentum.  We analyze this model following the procedure adopted by Stone~\cite{Stone}.  This analysis has been presented  in \cite{Deguchi-Fujikawa}, and thus we recapitulate the essence of the analysis. 

We start with the  Schr\"{o}dinger equation
$i\hbar \partial_{t}\psi(t)=H(t)\psi(t)$
or the Lagrangian
\begin{eqnarray}\label{generic Lagrangian}
L=\psi(t)^{\dagger}[i\hbar \partial_{t}-H(t)]\psi(t)
\end{eqnarray}
where the field $\psi(t)$ stands for the two-component spinor which describes the 
movement of two-levels crossing at the vanishing momentum.

We first perform a time dependent unitary transformation 
\begin{eqnarray}\label{unitary1}
\psi(t)= U(\vec{p}(t))\psi^{\prime}(t),\ \ 
\psi^{\dagger}(t)={\psi^{\prime}}^{\dagger}(t) 
U^{\dagger}(\vec{p}(t))
\end{eqnarray}
with
\begin{eqnarray}
U(\vec{p}(t))^{\dagger}\vec{p}(t)\cdot\vec{\sigma}
U(\vec{p}(t))
=|\vec{p}(t)|\sigma_{3}.
\end{eqnarray}
This unitary transformation is explicitly given by a $2\times2$ matrix
$U(\vec{p}(t))=\left(
             v_{+}(\vec{p})\  v_{-}(\vec{p}) \right)$,
where
\begin{eqnarray}\label{generic eigenfunction}
v_{+}(\vec{p})=\left(\begin{array}{c}
            \cos\frac{\theta}{2}e^{-i\varphi}\\
            \sin\frac{\theta}{2}
            \end{array}\right), \ \ \ 
v_{-}(\vec{p})=\left(\begin{array}{c}
            \sin\frac{\theta}{2}e^{-i\varphi}\\
            -\cos\frac{\theta}{2}
            \end{array}\right).
\end{eqnarray}
This unitary transformation corresponds to a use of instantaneous eigenstates of the operator $\mu\vec {p}(t)\cdot\vec{\sigma}$ where $\vec {p}(t)=|\vec{p}(t)|(\sin\theta\cos\varphi, \sin\theta\sin\varphi, \cos\theta)$.

Based on this transformation, 
 the equivalence of two 
Lagrangians is derived: $L$ in \eqref{generic Lagrangian}
and 
\begin{eqnarray}
L^{\prime}={{\psi}^{\prime}}^{\dagger}[i\hbar\partial_{t} +
\mu|\vec{p}(t)|\sigma_{3}+U(\vec{p}(t))^{\dagger}
i\hbar\partial_{t}U(\vec{p}(t))]\psi^{\prime}.
\end{eqnarray}
 The starting Hamiltonian \eqref{generic Hamiltonian} is thus replaced by  
\begin{eqnarray}\label{geometric-phase}
H^{\prime}(t)&=&-\mu|\vec{p}(t)|\sigma_{3}+ U(\vec{p}(t))^{\dagger}
\frac{\hbar}{i}\partial_{t}U(\vec{p}(t))\nonumber\\
&=& -\mu|\vec{p}(t)|\sigma_{3} -\hbar\left(\begin{array}{cc}
\frac{(1+\cos\theta)}{2}\dot{\varphi}&\frac{\dot{\varphi}\sin\theta
+i\dot{\theta}}{2}\\
            \frac{\dot{\varphi}\sin\theta
-i\dot{\theta}}{2}&
\frac{(1-\cos\theta)}{2}\dot{\varphi}
            \end{array}\right).
\end{eqnarray}
In the adiabatic approximation, 
\begin{eqnarray}
\mu|\vec{p}(t)| T\gg 2\pi\hbar,
\end{eqnarray}
where $T$ is the period of the dynamical
system $\vec{p}(t)$, and $2\pi\hbar$ stands for the magnitude of 
the geometric term times $T$, namely, we estimate $\dot{\varphi} \sim 2\pi/T$.
We then have 
\begin{eqnarray}\label{adiabatic Stone phase}
H^{\prime}_{ad} \simeq  -\mu|\vec{p}(t)|\sigma_{3} -\hbar\left(\begin{array}{cc}
\frac{(1+\cos\theta)}{2}\dot{\varphi}&0\\
            0&
\frac{(1-\cos\theta)}{2}\dot{\varphi}
            \end{array}\right)
\end{eqnarray}
since if $T$ is sufficiently large
one may neglect the off-diagonal parts in \eqref{geometric-phase} and retain only the diagonal components.

Stone~\cite{Stone} then finds
 that the adiabatic Berry's phase for the $++$ component
\begin{eqnarray}\label{adiabatic Stone phase2}
\exp[-i/\hbar \oint H^{\prime (++)}_{ad}dt]=\exp[i\mu\oint |\vec{p}(t)|/\hbar+i \oint \frac{(1+\cos\theta)}{2}d\varphi],
\end{eqnarray}
namely, the flux generated by a formal singularity located at the origin of the parameter space $\mu\vec{p}$ where two levels cross, 
\begin{eqnarray}\label{monopole-flux}
\Omega_{mono}=-\hbar\oint \frac{(1+\cos\theta)}{2}d\varphi
\end{eqnarray} 
is recognized as a monopole flux. In terms of the vector potential we have
\begin{eqnarray}
{\cal A}_{\varphi}=\frac{\hbar}{2|\vec{p}(t)|\sin\theta}(-1-\cos\theta)
\end{eqnarray}
in the lower hemisphere, which is gauge equivalent to 
\begin{eqnarray}\label{generic monopole}
{\cal A}_{\varphi}=\frac{\hbar}{2|\vec{p}(t)|\sin\theta}(1-\cos\theta)
\end{eqnarray}
  in the upper hemisphere with a Dirac string located at $\theta=\pi$, since the magnetic charge is given by $e_{M}=2\pi\hbar$
 as in \eqref{magnetic charge} and thus satisfies the Wu-Yang gauge invariance condition~\cite{Wu-Yang}. The adiabatic formula \eqref{generic monopole} agrees with the adiabatic limit in the solvable model \eqref{Dirac monopole}.

It is shown using the relation \eqref{geometric-phase} 
that if $\hbar$ times the frequency of $\vec{p}(t)$,
$2\pi\hbar/T$, is much larger than the level crossing energy  $\mu|\vec{p}(t)|$ or close to the level crossing point $|\vec{p}(t)|\rightarrow0$ with fixed $T$,  namely,
\begin{eqnarray}\label{non-adiabatic condition}
\mu|\vec{p}(t)| T\ll 2\pi\hbar,
\end{eqnarray}
then the geometric term dominates the $\mu|\vec{p}(t)|$ term. 
To see the implications of the condition \eqref{non-adiabatic condition} explicitly, one may perform a further (regular) unitary transformation of the 
fermionic  variable~\cite{fujikawa-mpla2005, Deguchi}
\begin{eqnarray}
\psi^{\prime}(t)= U(\theta(t))\psi^{\prime\prime}(t), \ \ \  
{\psi^{\prime}(t)}^{\dagger}=
{\psi^{\prime\prime}}^{\dagger}(t) 
U^{\dagger}(\theta(t))
\end{eqnarray}
with
\begin{eqnarray}\label{unitary2}
U(\theta(t))=\left(\begin{array}{cc}
            \cos\frac{\theta}{2}&-\sin\frac{\theta}{2}\\
            \sin\frac{\theta}{2}&\cos\frac{\theta}{2}
            \end{array}\right)
\end{eqnarray}
in addition to \eqref{unitary1}, which diagonalizes the dominant Berry phase term.
The Hamiltonian \eqref{geometric-phase} then becomes 
\begin{eqnarray}\label{Non-adiabatic Hamiltonian}
H^{\prime\prime}(t)
&=&
- U(\theta(t))^{\dagger}
\mu|\vec{p}(t)|\sigma_{3}U(\theta(t))
+ (U(\theta(t))U(\vec{n}(t)))^{\dagger}
\frac{\hbar}{i}\partial_{t}(U(\vec{n}(t))U(\theta(t)))\nonumber\\
&=&
-\mu|\vec{p}(t)| \left(\begin{array}{cc}
            \cos\theta&-\sin\theta\\
            -\sin\theta&-\cos\theta
            \end{array}\right)
-\hbar\left(\begin{array}{cc}
            \dot{\varphi}&0\\
            0&0
            \end{array}\right).
\end{eqnarray}
Note that the first term is bounded by $\mu|\vec{p}(t)|$ and the second term is dominant for  $\mu|\vec{p}(t)| T\ll 2\pi\hbar$. We emphasize that both \eqref{geometric-phase} and \eqref{Non-adiabatic Hamiltonian} are exact expressions. 

The Hamiltonian in the non-adiabatic approximation then becomes
\begin{eqnarray}\label{non-adiabatic Stone phase}
H^{\prime\prime}_{\rm nonad} \simeq
-\mu|\vec{p}(t)| \left(\begin{array}{cc}
            \cos\theta&0\\
            0&-\cos\theta
            \end{array}\right)
-\hbar\left(\begin{array}{cc}
            \dot{\varphi}&0\\
            0&0
            \end{array}\right).
\end{eqnarray}
The topological Berry's phase, which is independent of $\mu$ and $T$ after integration, thus either vanishes or becomes trivial independently of $\theta$
\begin{eqnarray}\label{trivial curvature}
\exp\{i\oint \dot{\varphi}dt\}=\exp\{2i\pi\}=1
\end{eqnarray}
for the very rapid movement $T\rightarrow0$ of $\vec{p}(t)$ or very close to the monopole position $|\vec{p}(t)|\rightarrow0$ with fixed $T$. 
Berry's phase  is thus topologically trivial (i.e., transformed to a trivial value under the 
continuous variation of the parameter $T\rightarrow 0$ with fixed $\mu|\vec{p}(t)|$ or $\mu|\vec{p}(t)|\rightarrow 0$ with fixed $T$)  and the monopole disappears $\oint\vec{{\cal A}}\cdot d\vec{p}=0$ up to $2\pi\hbar$~\cite{fujikawa-mpla2005, Deguchi}.   We emphasize that the non-adiabatic formula of Berry's phase \eqref{non-adiabatic Stone phase} agrees with the non-adiabatic limit of the exactly solvable 
model \eqref{solid-angle}.  In the two limiting cases, namely, at the adiabatic limit and the non-adiabatic limit, the solvable model \eqref{solid-angle} agrees with the present generic model.

We have demonstrated that the topology change  from the configuration with a Dirac monopole-like singularity to the topologically trivial configuration is smooth, in agreement with the analysis of an exactly solvable model \eqref{Hamiltonian}, and thus this behavior is generic. This smooth transition in the present generic model is facilitated by the regular transformation \eqref{unitary2}.  The transformation  \eqref{unitary2} may be called a resolution of monopole singularity in Berry's phase for the generic Hamiltonian \eqref{generic Hamiltonian} which is regular in the variable $\vec{p}(t)$ \cite{Deguchi-Fujikawa}.  We emphasize that all the precise formulas \eqref{generic Hamiltonian}, \eqref{geometric-phase} and \eqref{Non-adiabatic Hamiltonian} are unitary equivalent.

\section{Conclusion}
We have analyzed the basic properties of Berry's phase from a point of view of topology and topology change in a very explicit manner. 
We have identified a new kind of ``monopole'' in the sense that the geometrical center of a monopole-like configuration and the origin of the Dirac string, which appears  when the net outgoing flux is nonvanishing, are displaced in the parameter space as in Fig.5.
Gauss' theorem for a volume containing no singularity then shows that the basic topology of Berry's phase, which consists of a monopole-like configuration and a Dirac string, is always a dipole-like.  Only when the Dirac string satisfies the unobservability condition of Dirac~\cite{Dirac} and Wu and Yang~\cite{Wu-Yang}, a monopole-like object is identified, and otherwise we have a dipole-like object. We also mentioned the appearance of an interesting half-monopole with a magnetic charge $e_{M}/2$ and an observable Dirac string. We have thus revealed remarkably rich topological properties of the monopole-like object in Berry's phase and a novel mechanism of the smooth topology change from a monopole to a dipole when one approaches the monopole position in the parameter space. This smooth topology change is consistent with the resolution of monopole singularity in Berry's phase~\cite{Deguchi-Fujikawa}. The main topological features that are established by an exactly solvable model have been shown to be supported by a generic model of Berry's phase in section 3; this is expected since the topological properties are not very sensitive to the smooth deformation of various parameters.  

These explicit analyses should be useful to understand precisely what is the monopole-like object associated with Berry's phase.  The presence of the topology change shows that Berry's phase is different from the genuine Dirac monopole.
The present analyses are expected to be useful in the analysis of other basic aspects of Berry's phase, such as the consistency of adding Berry's phase to the canonical form of semi-classical equations of motion in condensed matter physics~\cite{Niu, Duval, Deguchi-Fujikawa}. These topological properties may  also  be useful in the analysis of the possible connection or no connection of Berry's phase with the notion of quantum anomalies~\cite{Stone, fujikawa-prd2006}.
\\

One of us (KF) is supported in part by JSPS KAKENHI (Grant No.18K03633).

\end{document}